\documentclass[12pt,dvips]{article}

\usepackage{amsmath,epsfig,ulem}
\usepackage{./url}
\usepackage{rotating}
\usepackage{cite}
\usepackage{hhline}
\usepackage{array}
\usepackage{axodraw}
\usepackage{amssymb}
\usepackage{color}

\begin{document}
\tolerance=100000

\newcommand{\imag}{\Im {\rm m}}
\newcommand{\real}{\Re {\rm e}}

\def\tablename{\bf Table}%
\def\figurename{\bf Figure}%

\newcommand{\sts}{\scriptstyle}
\newcommand{\ngs}{\!\!\!\!\!\!}
\newcommand{\rb}[2]{\raisebox{#1}[-#1]{#2}}
\newcommand{\CP}{${\cal CP}$~}
\newcommand{\sbomu}{\frac{\sin 2 \beta}{2 \mu}}
\newcommand{\kmol}{\frac{\kappa \mu}{\lambda}}
\newcommand{\s}{\\ \vspace*{-3.5mm}}
\newcommand{\lsim}{\raisebox{-0.13cm}{~\shortstack{$<$\\[-0.07cm] $\sim$}}~}
\newcommand{\gsim}{\raisebox{-0.13cm}{~\shortstack{$>$\\[-0.07cm] $\sim$}}~}
\newcommand{\kr}{\color{red}}
\newcommand{\RF}{${\mathbb{R} \!\!\! / \,\,\,}$}
\newcommand{\Rp}{${R_p \!\!\!\!\!\! / \,\,\,\,\,}$}

\begin{titlepage}

\begin{flushright}
DESY 06-238\\
IFT-06/027\\
MZ-TH/06-28\\
hep-ph/0612302
\end{flushright}

\vskip 1.5cm

\begin{center}
{\large \bf ISOLATED LEPTON EVENTS AT HERA:\\ SUSY \boldmath{$R$}-PARITY
  VIOLATION?}\\[0.9cm]
{\normalsize S.Y. Choi$^{1,2}$, J. Kalinowski$^{3}$,
   H.-U. Martyn$^{1,4}$, R. R\"{u}ckl$^{5}$, H. Spiesberger$^{6}$
     and P.M. Zerwas$^{1}$}\\[0.8cm]
{\it $^1$ Deutsches Elektronen-Synchrotron DESY, D-22603 Hamburg, Germany\\
     $^2$ Physics Department and RIPC, Chonbuk National University, Jeonju 561-756,
          Korea\\
     $^3$ Institute of Theoretical Physics, Warsaw University, PL-00681 Warsaw,
          Poland\\
     $^4$ I. Physikalisches Institut, RWTH Aachen, Aachen, Germany\\
     $^5$ Institut f\"{u}r Theoretische Physik, Universit\"{a}t W\"{u}rzburg,
          D-97074 W\"{u}rzburg, Germany\\

     $^6$ Institut f\"{u}r Physik, Johannes-Gutenberg-Universit\"{a}t
          Mainz, D-55099 Mainz, Germany.}
\end{center}

\renewcommand{\thefootnote}{\fnsymbol{footnote}}
\vspace{2.3cm}

\begin{abstract}
\noindent
Events with an isolated high $p_T$ lepton, a hadron jet and missing energy
as observed in the H1 experiment at HERA,
are potentially associated with $R$-parity violation in
supersymmetric theories. However, stringent
kinematic constraints must be fulfilled if the production of supersymmetric
particles in $R$-parity violating scenarios were the correct path for
explaining these puzzling events. A reference point \RF is specified for
which these constraints are illustrated and implications of the
supersymmetric interpretation for new classes of multi-lepton events are
indicated.
\end{abstract}

\end{titlepage}

\newpage
\renewcommand{\thefootnote}{\fnsymbol{footnote}}

\noindent
{\bf 1.} Several events with an isolated lepton in association with a hadron jet and
missing transverse momentum,
\begin{eqnarray}
e^+ p\ \ \to\ \ e^+/\mu^+\, +\, \mbox{jet}\, +\, p_T^{miss}
\label{eq:isolated_lepton_event}
\end{eqnarray}
have been observed in the H1 experiment at HERA. For events with the jet
transverse momentum larger than 25 GeV, the yield is larger, at a level of
$3.4\sigma$, than the number expected in the Standard Model \cite{H1}, in which
$W$ production \cite{Sp}, with subsequent leptonic decays, gives rise to these
final states. This type of events has been observed in running HERA with
positrons but not with electrons. ZEUS, on the other hand, has not reported the
observation of excess of such events \cite{ZEUS}. A direct
comparison of both experiments is difficult however since H1 and ZEUS
cover somewhat different phase-space regions \cite{Diaconu:2006qs}. \s

A potential interpretation of the events (\ref{eq:isolated_lepton_event}) is
based on supersymmetric theories incorporating $R$-parity violating
interactions [for a recent review of $R$-parity violating supersymmetry
see Ref.~\cite{R2}]. Most of the models assume the formation of fairly light
stop particles in $e^+d$ fusion, via the $R$-parity violating
interaction $\lambda'_{131}L_1Q_3D^c_1$ in the superpotential, with stops
decaying to a $b$-jet and a chargino.\footnote{Our discussion applies equally
  well to scenarios in which a charm squark is produced in  $e^+d$
  fusion via the $\lambda'_{121}$ coupling) and decays to a $s$-jet
  and a chargino.  However, due to possible strong mixing in the stop sector,
  the lightest stop can be expected considerably lighter than the charm squark.
  The models differ in the structure of the subsequent decay chains of
  the chargino.}\s

\noindent (i) In Refs.~\cite{H1,ZEUS} the chargino decays to a $W$-boson and
a neutralino $\tilde{\chi}^0_1$, with the $W$-boson
decaying leptonically. The final-state configuration in
Eq.$\,$(\ref{eq:isolated_lepton_event}), could be
achieved if the neutralino is assumed to be meta-stable, decaying outside the
detector. However, such an interpretation is very unlikely  since
the lifetime for $\tilde{\chi}^0_1$ decays is bounded from above by
the $\tilde{\chi}^0_1 \to b \tilde{b} \to b d \nu$ channel mediated
by  a virtual $\tilde{b}$. The size of the partial width, estimated as
$\Gamma(\tilde{\chi}^0_1)\sim 3 \alpha_W ({\lambda'}_{131}^2/4\pi)
m_{\tilde{\chi}^0_1}^5/32\pi m_{\tilde{b}}^4$,
is set by the size of the $e^+ d \tilde{t}$ coupling responsible for
the production process, see Eq.$\,$(\ref{eq:cycled}), and the
$\tilde{b}$ mass ($\alpha_W$ denotes the electroweak SU(2)$_I$ fine structure
constant $g^2_2/4\pi$). Similar consequences can be drawn when the
chargino decays to a lepton and a meta-stable sneutrino $\tilde{\nu}_\tau$
assumed to be the lightest SUSY particle, even if direct $R$-parity violating
$\tilde{\nu}_\tau$ couplings are absent, cf.~Ref.~\cite{sneut}.
The sneutrino may decay ``back'' through the channel $\tilde{\nu}_\tau\to
\tau\tilde{\chi}^+_1 \to \tau b \tilde{t} \to \tau b e^+d$ involving virtual
intermediate $\tilde{\chi}^+_1$ and $\tilde{t}$ states, with
$\Gamma(\tilde{\nu}_\tau)\sim  \alpha_W^2 ({\lambda'}^2_{131}/4\pi)
m^7_{\tilde{\nu}_\tau}/256\pi^2 m^2_{\tilde{\chi}^+_1}m^4_{\tilde{t}}$.
The estimated decay widths of the two modes, $\Gamma(\tilde{\chi}^0_1)
\sim 1 $ eV and $\Gamma( \tilde{\nu}_\tau) \sim 10^{-3}$ eV, for
$\lambda'_{131}\sim 5\times 10^{-2}$ and SUSY masses ${\cal O}$(100) GeV,
suggest lifetimes of order $\sim 10^{-15}$ sec and $\sim 10^{-12}$ sec,
respectively, much too short for the particles to escape the detector before
decay.\s

\noindent (ii) A different class of models is based on  \Rp
violation in both lepton-quark and lepton-lepton interactions. The
chargino, generated in the stop decay, decays to a charged slepton
and a neutrino, followed by the subsequent slepton \Rp decay to a
lepton$+$neutrino pair, cf. Ref.~\cite{R5,Diaconu:1998rf}. This
mechanism generates the characteristic final state of
Eq.$\,$(\ref{eq:isolated_lepton_event}) as shown in Fig.$\,$1(a).
The production of the intermediate stop particle is governed by the
\Rp term $\lambda'\,LQD^c$ in the superpotential, and the slepton
decay to a lepton $+$ neutrino pair by $\lambda\,LLE^c$. [The
slepton may also decay back, with strength $\lambda'$, to a top $+$
jet pair.] Moreover, if the chargino decays alternatively to a
sneutrino and a charged lepton, the final state, three charged
leptons and no missing transverse energy as predicted by the same
\Rp coupling, can be reconstructed {\it in toto}, Fig.$\,$1(b).
[Also in this case the lepton pair may be replaced by a jet pair.]\s

Besides the prediction of additional lepton and jet final states
in the second scenario, the clustering of partons and leptons
at invariant masses corresponding to the chargino mass in all events,
and in addition the sneutrino mass in multi-lepton events, provides
powerful internal cross-checks of the SUSY \Rp interpretation of any events
with signature (\ref{eq:isolated_lepton_event}).
[If the on-shell constraints fail, the escape to virtual intermediate
states is hardly a way out as rates fall dramatically in this case.]\s

The coupling $\lambda'$ must be sufficiently large, a few $10^{-2}$ for
$\lambda'_{131}$, to guarantee the necessary stop production rate.
This size can only be maintained if the suppression of rare lepton final
states in meson decays is reached {\it ad-hoc} either by
selection of non-vanishing ${\lambda}', \lambda$ couplings or
by fine-tuned destructive interference
between several $\lambda ', \lambda$ couplings. Though the fine-tuning of
order 10$^{-1}$ to 10$^{-2}$ remains moderate, both solutions
are based on {\it ad-hoc} assumptions  which superficially lack naturalness. \s

These general remarks will be illustrated in the analysis of a specific
reference point \RF in the next section. Many of the discussions will be
generic and the emphasis will not be on the detailed properties of the
model but on the general characteristics that can easily be transferred to
a wider scenario and to other scenarios. In particular the crucial clustering
conditions will
apply to any scenario based on stop production in the $s$-channel and they
are characteristic for a whole class of \Rp models for the supersymmetric
interpretation of the HERA events. The reference point \RF is chosen to
develop constraints on the  SUSY \Rp interpretation of experimental data but
it should not be mis-interpreted as an outstanding candidate for explaining
existing data. \s

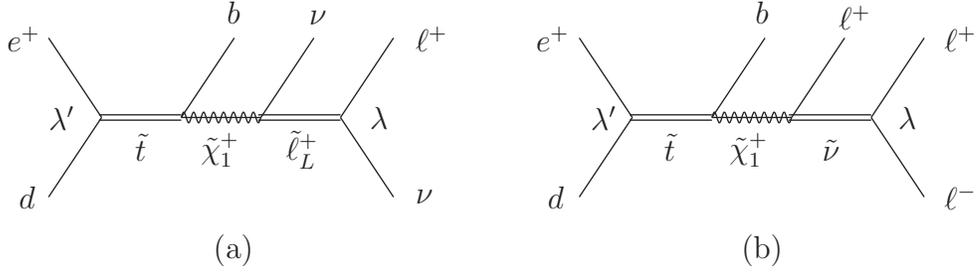
\begin{figure}[!h]
{
\begin{center}
\begin{picture}(400,110)(0,0)
\Text(27,80)[r]{$e^+$}
\Text(25,20)[r]{$d$}
\Line(30,80)(50,50)
\Line(30,20)(50,50)
\Text(35,50)[c]{$\lambda'$}
\Text(65,39)[c]{$\tilde{t}$}
\Line(49,51)(80,51)
\Line(49,49)(80,49)
\Text(100,90)[c]{$b$}
\Line(80,50)(100,80)
\Line(80,50)(110,50)
\Text(95,38)[c]{$\tilde{\chi}^+_1$}
\Photon(80,50)(110,50){2}{8}
\Text(132,88)[c]{$\nu$}
\Line(110,50)(130,80)
\Text(126,38)[c]{$\tilde{\ell}^+_L$}
\Line(109,51)(140,51)
\Line(109,49)(140,49)
\Line(140,50)(160,80)
\Line(140,50)(160,20)
\Text(169,80)[l]{$\ell^+$}
\Text(155,50)[c]{$\lambda$}
\Text(169,20)[l]{$\nu$}
\Text(100,0)[c]{(a)}
\Text(227,80)[r]{$e^+$}
\Text(225,20)[r]{$d$}
\Line(230,80)(250,50)
\Line(230,20)(250,50)
\Text(240,50)[c]{$\lambda'$}
\Text(265,39)[c]{$\tilde{t}$}
\Line(249,51)(280,51)
\Line(249,49)(280,49)
\Text(300,90)[c]{$b$}
\Line(280,50)(300,80)
\Line(280,50)(310,50)
\Text(295,38)[c]{$\tilde{\chi}^+_1$}
\Photon(280,50)(310,50){2}{8}
\Text(335,88)[c]{$\ell^+$}
\Line(310,50)(330,80)
\Text(326,38)[c]{$\tilde{\nu}$}
\Line(309,51)(340,51)
\Line(309,49)(340,49)
\Line(340,50)(360,80)
\Line(340,50)(360,20)
\Text(355,50)[c]{$\lambda$}
\Text(369,80)[l]{$\ell^+$}
\Text(369,20)[l]{$\ell^-$}
\Text(300,0)[c]{(b)}
\end{picture}
\end{center}
}
\caption{\it Supersymmetric $R$-parity violating interactions
generating isolated lepton events \label{fig:diagram3}}
\end{figure}

\vspace*{3mm}
\noindent
{\bf 2.} The reference point \RF is defined such that stop states are mixed
strongly and a rather light $\tilde{t}_1$ emerges while all the other
squark states are heavy. The light chargino and the lightest neutralino
are assumed higgsino-like so as to suppress $\tilde{\chi}^+_1 \to
W^+\tilde{\chi}^0_1$ decays in the $\tilde{t}_1$ cascade which, with
subsequent
\Rp violating $\tilde{\chi}^0_1$ decays, would not generate the desired
final states. The Higgs mixing parameter $\tan\beta$ is chosen moderate
to allow comparable charged slepton and sneutrino decays of the chargino.
[$R$-type sneutrinos are assumed to be very heavy and inaccessible.] \s

Constraints on the $\lambda'$ and $\lambda$ couplings in \RF from
rare mesonic decays may be illustrated by three important
examples~\cite{LElimits}:
\begin{eqnarray}
&& B^0_d \to \mu\mu\ \ : \ \
   |\lambda'_{131}\lambda^*_{122}+\lambda'_{331}\lambda^*_{322}|\ \ \lsim\ \
   4 \times 10^{-6}\, [\tilde{m}/100\,\mbox{GeV}]^2  \nonumber \\[1mm]
&& B^0_d \to \mu e\ \ \, : \ \
   |\lambda'_{131}\lambda^*_{121}+\lambda'_{331}\lambda^*_{321}|\ \ \lsim\ \
   8\times 10^{-6}\, [\tilde{m}/100\,\mbox{GeV}]^2\nonumber \\[1mm]
&& K^0_L \to \mu e\ \ : \ \
   |\lambda'_{1k1} {\lambda'}^*_{2k2}|\ \ \lsim\ \
   3\times 10^{-7}\, [\tilde{m}/100\,\mbox{GeV}]^2
   \label{eq:constraints}
\end{eqnarray}
where $\tilde{m}$ is the mass of the sparticle involved in the
decay. Vastly different $\lambda'$ and $\lambda$ values in any
product of the couplings or fine-tuning in destructive
interferences can suppress all these amplitudes. To illustrate the
first point, we may assume $\lambda'_{131}$ sufficiently large for
generating the necessary $\tilde{t}_1$ rate, and $\lambda_{322}$
or/and $\lambda_{321}$ sufficiently large for generating isolated $\mu$ and $e$
events [for the sake of simplicity assumed to be equal].
The constraints from rare decays can be complied  with in this configuration
if $\lambda'_{331}$ and $\lambda_{122, 121}$ are very small. The third condition
can be solved in a similar way or by destructive interferences in the amplitudes.
Even though such a specific choice may look unnatural, it cannot be refuted by
experimental results on the other hand. \s

In summary. The reference point \RF defined in Table 1 does not appear in
conflict with experimental data. The \Rp couplings of the superfields
generate interactions in the (s)quark and (s)lepton sectors of the type:
\newcommand{\ti}{\tilde}
\begin{eqnarray}
\lambda'_{131} L_1 Q_3 D^c_1 & \Rightarrow &
\lambda'_{131}\, ( e\,\tilde{t}\,d\,\, -\,\nu_e \tilde{b}\, d\,
              \,+\, {\rm  cycled}) \nonumber \\
\lambda_{322} L_3 L_2 E^c_2  &\Rightarrow &
\lambda_{322} \, ( \tau\tilde{\nu}_\mu \mu -\nu_\tau \tilde{\mu} \mu
              \,+\, {\rm cycled}) \nonumber \\
\lambda_{321} L_3 L_2 E^c_1 &\Rightarrow &
\lambda_{321} \, (\tau\,\tilde{\nu}_\mu\,  e -\nu_\tau \tilde{\mu} e
              \,+\, {\rm  cycled} )
\label{eq:cycled}
\end{eqnarray}
For brevity we use a simplified notation in which fields are supposed to
be cycled, with the tilde characterizing the sfermion field kept in the
middle position.\s

Masses and mixings generated by this reference point, and relevant for the
subsequent discussion, are listed in Table \ref{tab:tab2}.
These values are compatible with the bounds on masses and mixings
from LEP, Tevatron and HERA \cite{Bella}. If interpreted strictly within the
MSSM, the parameters would lead to too low a mass of the lightest Higgs
boson; however, this mass can be raised beyond the LEP limit in
extended theories \cite{King:2005my} without affecting the mass and mixing
parameters of Table~\ref{tab:tab2} significantly, cf. Ref.~\cite{Choi:2006fz}.\s

\begin{table}[htb]
\caption{\label{tab:tab1} {\it Definition of the reference point \RF} }
{
\begin{center}
\begin{tabular}{|l||l|}
\hline
\ \ \ \ \RF: Parameters\ \ \ \   &\ \ \ \  Values     \\
      \hline\hline
\ \ \ \    elw gaugino masses     &\ \ \ \  $M_2=2M_1=1.5$ TeV \\
\ \ \ \    higgsino mass          &\ \ \ \  $\mu=160$ GeV  \\
\ \ \ \    Higgs mixing           &\ \ \ \  $\tan\beta =1.5$ \\
       \hline
\ \ \ \    scalar lepton masses   &\ \ \ \  $M_L = M_E=130$ GeV \\
\ \ \ \    scalar quark masses    &\ \ \ \  $M_Q = M_U = M_D = 420$ GeV\\
\ \ \ \    trilinear $A$ coupling &\ \ \ \  $A_t = 840$ GeV \\
       \hline\hline
\ \ \ \    $\lambda', \lambda$ couplings
                  &\ \ \ \ $\lambda'_{131}=5\times 10^{-2}$ \\
\ \ \ \    { }    &\ \ \ \ $\lambda_{322}=\lambda_{321}=\lambda'_{131}$ \\
\ \ \ \    { }    &\ \ \ \ other $\lambda',\lambda$ very small\\
       \hline
\end{tabular}
\end{center}
}
\end{table}

\vskip -0.55cm
\begin{table}[htb]
\caption{\label{tab:tab2} {\it Masses and mixings at the reference point \RF}}
{
\begin{center}
\begin{tabular}{|l||l|}
\hline
\ \ \ \ \RF: Observables \ \ \ \   &\ \ \ \  Masses and mixing elements       \\
      \hline\hline
\ \ \ \    stop masses     &\ \ \ \  $m[\tilde{t}_{1/2}]=203/609$ GeV \\
\ \ \ \    stop mixing     &\ \ \ \  $[U_{\tilde{t}}]_{11}=-0.71$  \\
\ \ \ \    slepton/sneutrino masses
                           & \ \ \ \ $m[\tilde{\ell}_L/\tilde{\nu}]
                                      = 133/123$ GeV \\
       \hline
\ \ \ \    chargino masses   &\ \ \ \  $m[\tilde{\chi}^\pm_{1/2}]=156/1505$ GeV\\
\ \ \ \    chargino mixings  &\ \ \ \  $[U_{L/R}]_{11} = -0.049/-0.068$ \\
       \hline\hline
\ \ \ \    neutralino masses
                &\ \ \ \  $m[\tilde{\chi}^0_{1/2/3/4}]=152/160/753/1505$
                                        GeV \\
\ \ \ \    neutralino mixings
                &\ \ \ \  $[N]_{11/2/3/4}=-0.070/0.058/-0.71/0.70 $ \\
       \hline
\end{tabular}
\end{center}
}
\end{table}

\noindent
{\bf 3.} The production cross section of stop particles at HERA,
cf. Figs.$\,$1, is given in the narrow-width approximation by
\begin{eqnarray}
&& \sigma(e^+p\to\tilde{t}_1)\,\,\ \ = \ \ \hat{\sigma}_0(e^+d\to\tilde{t}_1)\,
                                         \, f_d (m^2_{\tilde{t}_1}/s) \\
&& \hat{\sigma}_0(e^+d\to\tilde{t}_1) \ \ = \ \
   \frac{\pi}{4 m^2_{\tilde{t}_1}} |
   \, \lambda'_{131}|^2\, |[U_{\tilde{t}}]_{11}|^2\nonumber
\end{eqnarray}
with the $d$-parton density $f_d$ in the proton evaluated at
$x=m^2_{\tilde{t}_1}/s$. For the \RF parameter set introduced above and
the MRST2004 parametrization of the parton densities \cite{msrt2004},
the
numerical value of $\sigma(e^+p\to\tilde{t}_1)=0.54$ pb leads to a
sample of about ninety $\tilde{t}_1$
events for an integrated luminosity of $\int\! {\cal L} =160$ pb$^{-1}$.\s

Since the $t\tilde{\chi}^0_1$ decay channel is kinematically inaccessible,
$\tilde{t}_1$ particles decay to nearly 100\% into $b\tilde{\chi}^+_1$ final
states, as indicated in Figs.$\,$1; remnant $\not\!\! R_p$ decays to
$e^+/\tau^+ d$ are reduced to a level of 3\% as a result of the small
$\lambda'_{131/331}$ couplings compared with the gauge coupling.\s

Also the subsequent chargino $\tilde{\chi}^+_1$ decays follow the standard
path to $\tilde{\ell}^+_L\nu_\ell$ and $\tilde{\nu}\ell^+$ final states since
the chargino decay to $W^+\tilde{\chi}^0_1$ is strongly suppressed
$\sim (m_{\tilde{\chi}^+_1}-m_{\tilde{\chi}^0_1})^5$ in the higgsino
region with nearly degenerate $\tilde{\chi}_1^+/\tilde{\chi}_1^0$ states,
i.e. $\Delta m_{\tilde{\chi}}
\simeq 4$ GeV in \RF\!\!. The partial widths are given by
\begin{eqnarray}
\Gamma(\tilde{\chi}^+_1\to\tilde{\ell}^+_{L}\nu_\ell/\tilde{\nu}_\ell \ell^+)
 = \frac{\alpha}{16 \sin^2 \theta_W}\, m_{\tilde{\chi}^\pm_1}\,
   \left[1-m^2_{\tilde{\ell}_L/\tilde{\nu}}/m^2_{\tilde{\chi}^\pm_1}
   \right]^2\, |[U_{L/R}]_{11}|^2
\end{eqnarray}
To generate a sufficiently large number of isolated lepton plus missing $p_T$
events, the Higgs mixing parameter $\tan\beta$ cannot be much larger than
unity; this
is required by the electroweak $D$ term $m^2_{\tilde{\nu}}-m^2_{\tilde{\ell}_L}
=m^2_W \cos 2\beta$ as well as the ratio of mixing elements
$[U_L]_{11}/[U_R]_{11}\sim 1/\tan\beta$. The branching ratio for charged
slepton
final states amounts to 21\% so that a significant number of sneutrino final
states is predicted in the \RF environment. Reducing $\tan\beta$
further leads to near balance between charged and neutral final states.\s

For the decays of the charged sleptons and sneutrinos a variety of channels
may be open. In addition to the leptonic $\not\!\! R_p$ decays, hadronic
$\not\!\! R_p$ decays and standard gaugino decays to $\tilde{\chi}^0_1$
may be observed, with $\tilde{\chi}^0_1$ finally decaying to leptons and
hadrons through $\not\!\! R_p$ interactions mediated by a virtual $\tilde{b}$
state:
\begin{eqnarray}
\tilde{\ell}^+_L  &\to &  e^+ \nu_\mu,\, e^+ \nu_\tau,\, \mu^+ \nu_\mu,\,
                          \mu^+ \nu_\tau
                         \, \,\,   ;\,
                      t \bar{b}\,\,;\, \ell^+ \tilde{\chi}^0_1 \nonumber\\
\tilde{\nu}_\ell  &\to & e^-\mu^+, e^-\tau^+, \mu^-\mu^+, \mu^-\tau^+\,;\,
                      d\bar{b}\,;\, \nu_\ell \tilde{\chi}^0_1
\end{eqnarray}
The branching ratios depend strongly on the details of the parameter
choice, apparent from the partial widths:
\begin{eqnarray}
 \Gamma(\tilde{\ell}_L\to \ell\nu/ \tilde{\nu}\to\ell\ell')
    \,& =&\, \frac{|\lambda|^2}{16\pi}\, m_{\tilde{\ell}_L/ \tilde{\nu}} \\[1mm]
 \Gamma(\tilde{\ell}_L\to td/ \tilde{\nu}\to bd)
    \,& =&\,  \frac{3|\lambda'|^2}{16\pi} m_{\tilde{\ell}_L/ \tilde{\nu}}
      \left[1-{m^2_{t/b}}/{m^2_{\tilde{\ell}_L/\tilde{\nu}}}\right]^2\\
 \Gamma(\tilde{\ell}_L\to \ell\tilde{\chi}^0_1/ \tilde{\nu}\to
\nu\tilde{\chi}^0_1)
    \, &=&\,  \frac{\alpha}{4 \sin^2 2 \theta_W} |N_{12} c_W + N_{11} s_W|^2
      m_{\tilde{\ell}_L/\tilde{\nu}}
   \left[1-{m^2_{\tilde{\chi}^0_1}}/{m^2_{\tilde{\ell}_L/\tilde{\nu}}}\right]^2
\end{eqnarray}
In the reference scenario selectrons and $e$-sneutrinos decay into
hadrons, but the other $\mu, \tau$ type sleptons decay to leptons.
Standard neutralino decays are also suppressed due to the reduced phase-space
if the neutralino mass is close to the slepton/sneutrino
masses and if the neutralino is higgsino-like.\s

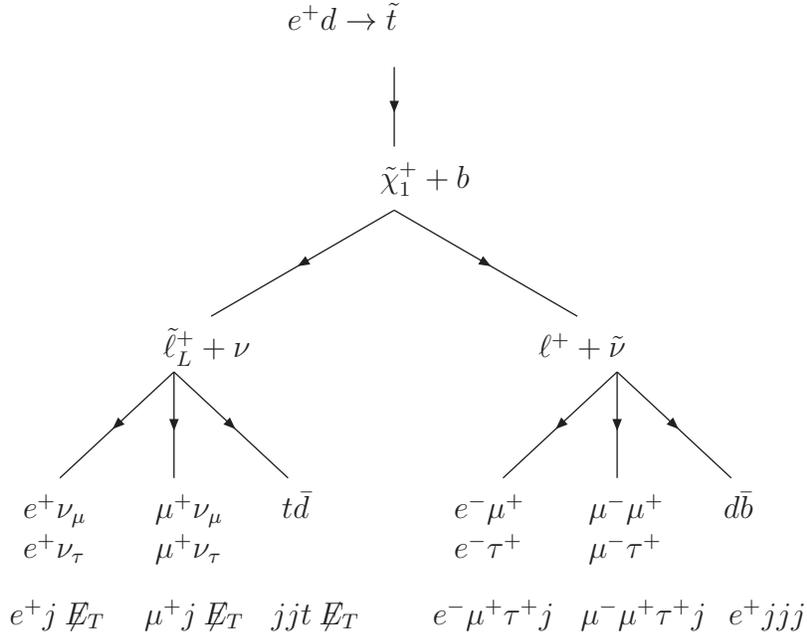
\begin{figure}[htbp]
\begin{picture}(200,240)(-125,0)
\put(85,235){$e^+d \rightarrow \tilde{t}$}
\ArrowLine(125,220)(125,190)
\put(120,175){$\tilde{\chi}_1^+ + b$}
\ArrowLine(125,166)(56,126)
\ArrowLine(125,166)(194,126)
\put(38,111){$\tilde{\ell}^+_L + \nu$}
\put(180,111){$\ell^+ + \tilde{\nu}$}
\ArrowLine(42,105)(-1,65)
\ArrowLine(42,105)(42,65)
\ArrowLine(42,105)(85,65)
\ArrowLine(209,105)(166,65)
\ArrowLine(209,105)(209,65)
\ArrowLine(209,105)(252,65)
\put(-15,50){$e^+\nu_\mu$}
\put(-15,35){$e^+\nu_\tau$}
\put(35,50){$\mu^+\nu_\mu$}
\put(35,35){$\mu^+\nu_\tau$}
\put(83,50){$t\bar{d}$}
\put(148,50){$e^-\mu^+$}
\put(199,50){$\mu^-\mu^+$}
\put(250,50){$d\bar{b}$}
\put(148,35){$e^-\tau^+$}
\put(199,35){$\mu^-\tau^+$}
\put(-20,10){$e^+ j\not\!\! E_T$}
\put(31,10){$\mu^+ j\not\!\! E_T$}
\put(79,10){$jjt\not\!\! E_T$}
\put(140,10){$e^-\mu^+\tau^+  j$}
\put(196,10){$\mu^-\mu^+\tau^+  j$}
\put(252,10){$e^+ jjj$}
\end{picture}
\caption{\label{fig:tree} \it Mixed $R$-parity conserving and $R$-parity
  violating decays of the lighter top squark $\tilde{t}_1$ which give
  rise to multi-lepton and jet final states, in the left cascade including
  missing transverse momentum due to escaping neutrinos. }
\end{figure}

The chargino decay modes in the \RF scenario are summarized in Table 3,
leading to the experimentally observable final states indicated in the two
branches of Fig.$\,$\ref{fig:tree}
and for charged leptons and sneutrinos. All multi-lepton events in
\RF
contain $\tau^+$'s in the final state.  \s

\begin{table}[htb]
\caption{\label{tab:tab3} {\it The chargino and the subsequent slepton/sneutrino
decay modes and the experimentally observable final states in the \RF scenario.
Estimates of the combined branching ratios for the
${\tilde{t}}_1$ decays to the observable final states are listed in the last column.
[The top quark in $\tilde{e}^+_L\to t\bar{d}$ is virtual as
$m_{\tilde{e}_L}< m_t$.] }}
{
\begin{center}
\begin{tabular}{|c|ccc||c|c|}
\hline
\multicolumn{6}{|c|} {$\tilde{t}_1\to b\tilde{\chi}^+_1\to b\tilde{\ell}^+_L\nu$
  and
    $b\tilde{\nu}\ell^+$ } \\
    \hline
$\tilde{\chi}^+_1$ decay & $\lambda_{322}$     & $\lambda_{321}$
                         & $\lambda'_{131}$    & $e^+p\to\tilde{t}_1$ final
                         state & Estimated fractions (\%)\\
       \hline\hline
$\tilde{e}^+_L\nu_e$     & $-$                 & $-$
                         & $t^*\bar{d}$          & $jjW^+b \not\!\! E_T$
                         & 7.0 \\
$\tilde{\mu}^+_L\nu_\mu$     & $\mu^+\nu_\tau$ & $e^+\nu_\tau$
                         & $-$         & $\mu^+ j \!\not\!\! E_T$ and
                                         $e^+ j\!\not\!\! E_T$
                         & 3.5 and 3.5 \\
$\tilde{\tau}^+_L\nu_\tau$   & $\mu^+\nu_\mu$  & $e^+\nu_\mu$
                         & $-$         & $\mu^+ j \!\not\!\! E_T$ and
                                         $e^+ j \!\not\!\! E_T$
                         & 3.5 and 3.5 \\
       \hline
$\tilde{\nu}_e e^+$      & $-$                & $-$
                         & ${d}\bar{b}$       & $e^+ jjj $
                         & 23.0 \\
$\tilde{\nu}_\mu \mu^+$    & $\mu^-\tau^+$    & $e^-\tau^+$
                         & $-$        & $\mu^-\mu^+\tau^+ j$ and $e^-\mu^+\tau^+j$
                         & 11.5 and 11.5 \\
$\tilde{\nu}_\tau \tau^+$    & $\mu^-\mu^+$      & $e^-\mu^+$
                         & $-$        & $\mu^-\mu^+\tau^+ j$ and $e^-\mu^+\tau^+ j$
                         & 11.5 and 11.5 \\
       \hline
\end{tabular}
\end{center}
}
\end{table}

A large variety of experimental signatures emerges from the model, with
no overwhelming preference for a particular channel, but large uncertainties
due to the
choice of mass, mixing and $\lambda', \lambda$ parameters. The table should
therefore not be interpreted as a quantitative prediction but rather as
a listing of interesting signatures to be searched for experimentally.\s

\noindent
{\bf 4.} In on-shell $\tilde{t}_1$ production in $e^+d$ collisions,
         the $\tilde{t}_1$ mass determines the  value $x$ of the target
         $d$ parton,
\begin{eqnarray}
x =m^2_{\tilde{t}_1}/s
\end{eqnarray}
The 4-momentum of the stop quark, traveling along the proton
direction,
is predicted to be
\begin{eqnarray}
p_{\tilde{t}_1}=p_e+(m^2_{\tilde{t}_1}/s) p_p
\end{eqnarray}
in obvious notation for the momenta of $\tilde{t}_1$, positron and proton.
Energy and 3-momentum of $\tilde{t}_1$ along the proton axis are fixed
in the laboratory frame by
\begin{eqnarray}
E_{\tilde{t}_1}   &=& \phantom{+} E_e + (m^2_{\tilde{t}_1}/s) E_p\nonumber\\
p^z_{\tilde{t}_1} &=& -E_e + (m^2_{\tilde{t}_1}/s) E_p
\label{eq:tilde_t_momentum}
\end{eqnarray}
These relations for the stop energy and momentum serve as basis for
evaluating the kinematical constraint equations which follow from the on-shell condition
for $\tilde{\chi}^+_1$. \s

In all scenarios discussed above, and universally for all SUSY \Rp
scenarios analyzed in the literature we are aware of, the decay of
$\tilde{t}_1$ always proceeds through $b\, \tilde{\chi}^+_1$ if the
final state in Eq.(\ref{eq:isolated_lepton_event}) is generated.
Using the $\tilde{t}_1$ 4-momentum from
Eq.$\,$(\ref{eq:tilde_t_momentum}) and the measured  $b$-jet energy
$E_b$ and longitudinal $z$-momentum $p_b^z$ along the proton
direction, the $\tilde{\chi}^+_1$ mass condition
$m^2_{\tilde{\chi}^+_1}=(p_{\tilde{t}_1}-p_b)^2$ can be cast in the
form
\begin{eqnarray}
m^2_{\tilde{\chi}^\pm_1} \ \ = \ \
m^2_{\tilde{t}_1}[1 - (E_b-p_b^z)/2E_e] -2 E_e(E_b+p^z_b) \label{eq:onshell}
\end{eqnarray}
With the {\it a priori} unknown stop and chargino masses,
each event with measured values of
$b$-jet energy and longitudinal momentum defines,
according to Eq.$\,$(\ref{eq:onshell}), a line in the
(mass)$^2$ plane, the coordinates labeled by $(m^2_x,m^2_y)$. \s

\begin{figure}[htb!]
\begin{center}
\includegraphics*[height=9.8cm,width=16.7cm,angle=0]{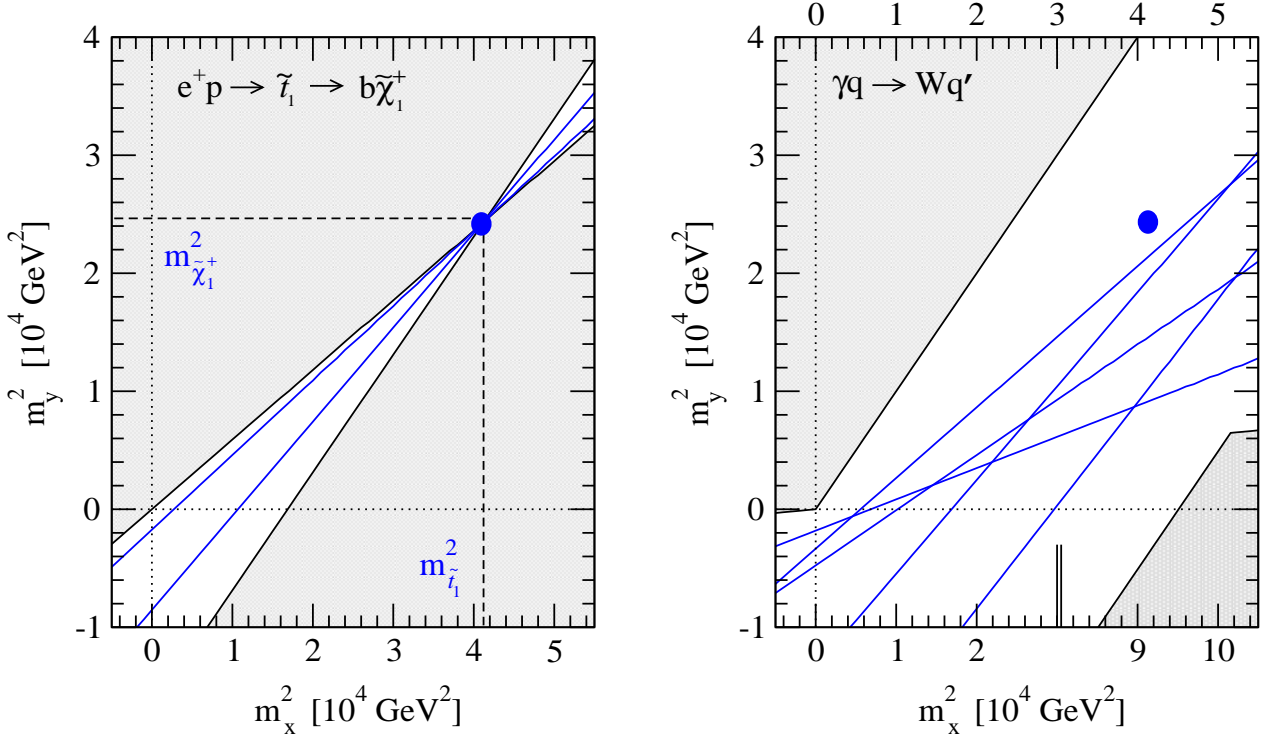}
\end{center}
\caption{\it Left: Lines corresponding to \Rp signal events.
 All signal lines  must lie within the double cone.
 The (blue) dot corresponds to the signal point for the true mass values,
 $m^2_x \to m^2_{\tilde{t}_1}$ and $m^2_y \to m^2_{\tilde{\chi}^\pm_1}$.
 Right: Lines corresponding to background events from $W$-photoproduction
 $\gamma q\to Wq' $. The kinematically excluded regions bounded by black solid
 lines are shaded. [Note that the lower $x$-axis is cut to accommodate the
 forbidden wedge on the right.]}
\label{fig:xcross}
\end{figure}

Lines corresponding to the signal events must cross at the single point
corresponding to the true values of stop and chargino masses
$(m^2_x,m^2_y)=(m^2_{\tilde{t}_1},m^2_{\tilde{\chi}^\pm_1})$, if the
considered \Rp scenario is the correct interpretation of the data.
This is demonstrated in Fig.\,\ref{fig:xcross}(a) for the reference
point \RF\!\!. All allowed signal lines in the mass plane are
located within the double cone formed by the line with slope = 1 and
cutting the $m^2_y$-axis at the minimum value = $-(m^2_{\tilde{t}_1}-
m^2_{\tilde{\chi}^\pm_1})$, and the line crossing the origin with
slope = $m^2_{\tilde{\chi}^\pm_1}/m^2_{\tilde{t}_1}$, both intercept and
slope given by the true mass values. The opening angle of the
cone increases apparently with the $\tilde{t}_1 - \tilde{\chi}^\pm_1$
mass gap. \s

On the other hand, lines corresponding to background events do not
cross at a single point but rather form an irregular mesh over the
$(m^2_x,m^2_y)$ plane. This is shown in Fig.$\,$\ref{fig:xcross}(b) for a
few examples for which the jet parameters are identified with values
derived from $W$-photoproduction $\gamma q\to Wq'$. Inserting jet
energy and momentum for a sample of events apparently does not lead
to any clustering
of the solutions (in contrast to the signal lines), reflecting that
the process does not proceed through resonance formation. For
$W$-photoproduction the lines fill the area between (and including)
the double-cone of maximum  $\gamma q$ invariant  mass, $M^2_{\gamma q}=s$,
and the line corresponding to the degenerate double-cone for minimum
$\gamma q$ invariant  mass, $M^2_{\gamma q}=M^2_W$. \s\s

Apparent from Figs.$\,$1(a) and (b) and Table \ref{tab:tab3}, many more
consistency conditions
can be derived from the leptons and jets in the final state.\s

Two examples are provided by the first and fourth row in  Table \ref{tab:tab3}.
The $\tilde{e}^+_L$ events in the first row must cluster at the invariant mass
\begin{eqnarray}
M[jt]             & = & m_{\tilde{e}_L}
\end{eqnarray}
while the $\tilde{\nu}_e$ events in the fourth row must cluster at
the triple point
\begin{eqnarray}
M[jj]             & = & m_{\tilde{\nu}_e} \nonumber \\
M[e^+ jj]         & = & m_{\tilde{\chi}^+_1}\nonumber \\
M[e^+ jjj]        & = & m_{\tilde{t}_1}
\end{eqnarray}

Moreover, since the $\nu_\tau$ 3-momentum can be reconstructed fully in
single $\tau$ events, the last two rows of Table 3 accounting for
all sneutrino decays, must fulfill a large set of constraint
equations. This may be illustrated by analyzing the $\mu\mu$ chain of
the last row in Table 3:
\begin{eqnarray}
M[\mu\mu]         & = & m_{\tilde{\nu}_\tau} \nonumber \\
M[\tau^+\mu\mu]   & = & m_{\tilde{\chi}^+_1}\nonumber \\
M[j\tau^+\mu\mu]  & = & m_{\tilde{t}_1}
\end{eqnarray}
and analogously for the other chains. Particularly appealing is the
requirement of clustering for the 2-lepton and 3-lepton invariant
masses at the $\tilde{\nu}$ and $\tilde{\chi}^+_1$ masses.\s

It should also be noted that the transverse energies of the $b$-jets in the
laboratory cluster for
$\tilde{t}_1 \to b \tilde{\chi}^+_1$ decays at the maximum possible
value
\begin{equation}
max\,[E_T^b]  = (m^2_{\tilde{t}_1} - m^2_{\tilde{\chi}^+_1})/2m_{\tilde{t}_1}
\end{equation}
as a result of the well-known transverse Jacobian
peak. \s

\vspace*{3mm}
\noindent
{\bf 5.} In this brief note we have examined necessary requirements which the
interpretation of isolated lepton events with large missing transverse momentum
and a hadron jet at HERA must comply with in the context of $R$-parity violating
supersymmetric interactions. Independent of the specific reference point used in
this note merely to illustrate the analysis in detail, two generic implications
have emerged from the study:
\begin{itemize}
\item[{--}] Kinematical constraint equations relate the observed jet and lepton
      energies and momenta with masses of stop, chargino and sneutrino, and
      clusters of invariant  masses must be observed experimentally.
\item[{--}] $R$-parity violating couplings connect leptons and sleptons of different
      charges and species. This implies that also events including multi-lepton
      final states and $\tau^+$ leptons should be observed.
\end{itemize}
The lifetime of meta-stable neutralinos and sneutrinos
is estimated to be very likely  too short to provide
a possible alternative to the generic type of scenarios analyzed above.\s

\subsubsection*{Acknowledgments}

We should like to thank G.~Brandt for providing us with the
$W$-photoproduction events. A communication on the Higgs sector by
F.~Richard is gratefully acknowledged.
The work was supported in part by the Korea Research Foundation
Grant (KRF-2006-013-C00097), by KOSEF through CHEP at Kyungpook
National University, by the Polish Ministry of Science and
Higher Education Grant No~1~P03B~108~30,
and by the German Ministry of Education and Research (BMBF) under
contract number 05HT6WWA. S.Y.C. is grateful for
support during his visit to DESY.



\begin{thebibliography}{99}
%
\bibitem{H1}
  A.~Aktas {\it et al.}  [H1 Collaboration],
  Eur.\ Phys.\ J.\ C {\bf 36} (2004) 425
  [hep-ex/0403027];
  D. South, {\it Proceedings, DIS2006, Tsukuba}, 2006,
  \url{http://www-conf.kek.jp/dis06/doc/WG3/ew06-south.pdf};
  H1 Collaboration, {\it Contributed paper to ICHEP, Moscow}, 2006,
  \url{http://www-h1.desy.de/h1/www/publications/htmlsplit/
  H1prelim-06-162.long.html}.

\bibitem{Sp}
  K.~P.~Diener, C.~Schwanenberger and M.~Spira,
  Eur.\ Phys.\ J.\ C {\bf 25} (2002) 405
  [hep-ph/0203269];
  P.~Nason, R.~R\"uckl and M.~Spira,
  J.\ Phys.\ G {\bf 25} (1999) 1434
  [hep-ph/9902296].

\bibitem{ZEUS} ZEUS Collaboration,
  {\it Contributed paper to ICHEP, Moscow}, 2006,
  \url{http://www-zeus.desy.de/physics/phch/conf/ichep06}, ZEUS-prel-06-012.

\bibitem{Diaconu:2006qs}
  C.~Diaconu, {\it Talk at ICHEP2006, Moscow},
  arXiv:hep-ex/0610041.

\bibitem{R2} R.~Barbier {\it et al.},
  Phys.\ Rept.\  {\bf 420} (2005) 1
  [hep-ph/0406039].

\bibitem{sneut} D. South, {\it Talk at DIS2006, Tsukuba},
   \url{http://www-conf.kek.jp/dis06/transparen cies/WG3/ew-south.pdf};
   E.~Sauvan, {\it Talk at DIS2006, Tsukuba},
   \url{http://www-conf.kek.jp/dis06/transparencies/summary/ew/pl-sauvan.pdf}.

\bibitem{R5} J.~Kalinowski, R.~R\"uckl, H.~Spiesberger and P.M.~Zerwas,
   DESY Internal Note [unpublished], 1997.

\bibitem{Diaconu:1998rf}
  C.~Diaconu, J.~Kalinowski, T.~Matsushita, H.~Spiesberger and D.~S.~Waters,
  {\it Proceedings, 3rd UK Workshop on HERA Physics, Durham},
  J.\ Phys.\ G {\bf 25} (1999) 1412
  [hep-ph/9901335].

\bibitem{LElimits}
  J.~H.~Jang, J.~K.~Kim and J.~S.~Lee,
  Phys.\ Rev.\ D {\bf 55} (1997) 7296
  [hep-ph/9701283];
  H.~K.~Dreiner, G.~Polesello and M.~Thormeier,
  Phys.\ Rev.\ D {\bf 65} (2002) 115006
  [hep-ph/0112228];
  R.~Van Kooten,
  {\it Proceedings of 4th Flavor Physics and CP Violation
  Conference (FPCP 2006), Vancouver BC},
  arXiv:hep-ex/0606005.

\bibitem{Bella}
  L.~Bellagamba,
  {\it Stop production in R-parity violating supersymmetry at Tevatron},
  arXiv:hep-ex/0611012.

\bibitem{King:2005my}
  S.~F.~King, S.~Moretti and R.~Nevzorov,
  Phys.\ Lett.\ B {\bf 634} (2006) 278
  [hep-ph/0511256].

\bibitem{Choi:2006fz}
  S.~Y.~Choi, H.~E.~Haber, J.~Kalinowski and P.~M.~Zerwas,
  {\it The neutralino sector in the U(1)-extended supersymmetric standard model},
  arXiv:hep-ph/0612218.

\bibitem{msrt2004} A.~D.~Martin, W.~J.~Stirling and R.~S.~Thorne, {\it MRST
   partons generated in a fixed-flavor scheme}, arXiv:hep-ph/0603143.

\end{thebibliography}
\end{document}